\documentclass[12pt]{article}
\usepackage{epsfig}
\usepackage{amssymb,amsmath,amsfonts}
\usepackage{MnSymbol,pifont,color}

\def\beq{\begin{equation}}
\def\eeq#1{\label{#1}\end{equation}}
\def\eeqn{\end{equation}}

\def\beqa{\begin{eqnarray}}
\def\eeqa#1{\label{#1}\end{eqnarray}}
\def\eeqan{\end{eqnarray}}

\let\bar=\overbar

\def\Dslash{\not{\hbox{\kern-4pt $D$}}}
\def\dslash{\not{\hbox{\kern-2pt $\del$}}}

\def\msb{{\bar{\ssstyle M \kern -1pt S}}}

\def\Title#1{\begin{center} {\Large {\bf #1} } \end{center}}

\begin{document}

\hfill SI-HEP-2012-20, QFET-2012-01, EOS-2012-03

\bigskip

\Title{Implications of the experimental results on 
       rare $b\to s + (\gamma,\,\ell^+\ell^-)$ decays}

\bigskip


%
%

\begin{raggedright}  

{\it Frederik Beaujean$^{\,a)}$, Christoph Bobeth$^{\,b)}$\footnote{Speaker} and 
     Danny van Dyk$^{\,c)}$\\[0.4cm]
$^{\,a)}$ 
  Max-Planck-Institut f\"ur Physik, F\"ohringer Ring 6, 80805 M\"unchen, Germany\\[0.1cm]
$^{\,b)}$ 
  Technische Universit\"at M\"unchen, Universe Cluster, 85748 Garching, Germany\\[0.1cm]
$^{\,c)}$ 
  Theoretische Physik 1, Naturwissenschaftlich-Technische Fakult\"at, \\
  \hskip 0.5cm Universit\"at Siegen, Walter-Flex-Stra\ss{}e 3, 57068 Siegen, Germany}

\end{raggedright}

\bigskip

{\it
  \noindent 
  Proceedings of CKM 2012, the 7th International Workshop on the CKM
  Unitarity Triangle, University of Cincinnati, USA, 28 September - 2
  October 2012
}

\begin{abstract}
The experimental measurements of flavor-changing neutral-current $B$-meson 
decays governed by $b\to s + (\gamma,\,\ell^+\ell^-)$ transitions have
entered a new level of precision. Recent results by Belle, CDF, Babar, 
and LHCb on $B\to K^{(*)}\ell^+\ell^-$ and $B_s \to \mu^+ \mu^-$ decays
are used in model-(in)dependent analyses to test the Standard Model
predictions and to derive stronger constraints on nonstandard contributions.
While in agreement with the Standard Model, they still leave sizable room for
new physics. 
 
\end{abstract}

%
%
%

Flavor-changing neutral-current (FCNC) decays of $B$ mesons, mediated by the
transitions $b\to s + (\gamma,\,\ell^+\ell^-)$, are probed experimentally
with unprecedented precision. The results of Belle \cite{Wei:2009zv}, CDF 
\cite{Aaltonen:2011qs} and BaBar \cite{:2012vwa} with about 200 events for 
$B\to K^{(*)}\ell^+\ell^-$ ($\ell = e,\,\mu$) are currently supplemented with
LHCb measurements \cite{Aaij:2012cq}, based on 1 fb$^{-1}$ from 2011 with about
1000 events. Whereas BaBar and CDF have already analyzed their final data sets,
the Belle results are from a partial data set only. The 2012 data set of LHCb 
with $\gtrsim 2$ fb$^{-1}$ will hopefully allow first measurements of angular
observables in $B\to K^{*}\ell^+\ell^-$ and the precision will improve further
with data of about $(3-4)$ fb$^{-1}$ in the years 2015 - 2018, before a shutdown 
for the planned upgrade. With the start of data taking in 2015, also the super-flavor
factory Belle II will collect a substantial data set with $(1.0 - 1.5) \cdot 10^4$ 
events \cite{Bevan:2011br} in the next decade. Very recently, LHCb found first
evidence for the very rare decay $B_s\to \mu^+\mu^-$ \cite{:2012ct}, known 
as an ideal probe of scalar and pseudo-scalar nonstandard interactions.


Currently, the measured observables in the exclusive channels $B\to K^*\gamma$,
$B\to K^{(*)}\ell^+\ell^-$ and $B_s \to \mu^+\mu^-$ comprise branching ratios 
(${\cal B}$), lepton forward-backward asymmetries ($A_{\rm FB}$), longitudinal
$K^*$-polarization fraction ($F_L$), the flat term ($F_H$), the angular 
observables $A_T^{(2)},\, S_3$ and $A_{\rm im},\,S_9$, isospin asymmetries ($A_I$) and
rate CP asymmetries ($A_{\rm CP}$)  in various bins of the 
dilepton invariant mass $q^2$ as well as the mixing-induced ($S$) and 
direct ($C$) CP asymmetry in $B\to K^*\gamma$. 

Theory predictions for $B\to K^{(*)} \ell^+\ell^-$ focus on the two regions in $q^2$ 
above and below the two narrowly peaking $c\bar{c}$-resonances $J/\psi$ and $\psi'$. 
At low $q^2$, the large recoil energy of the light meson allows to apply QCD 
factorization \cite{Beneke:2001at} and resonances can be included with the help 
of a non-local OPE in combination with dispersion relations~\cite{Khodjamirian:2010vf}.
At high $q^2$, nonfactorizable contributions are treated within a local OPE
\cite{Grinstein:2004vb}. Consequently, in the studies, only measurements in
$q^2$-bins are used that reside in these regions, i.e., $q^2 \lesssim (6-7)$~GeV$^2$
and $q^2 \gtrsim (14 - 15)$~GeV$^2$, respectively. A smaller $q^2$ binning of
future data will allow to benefit from the spectral information, as for example
positions of the zero crossings or maxima and minima.

Form factors are a crucial ingredient for observables like ${\cal B}$, $A_{\rm FB}$, 
$F_L$ and form the bulk of theoretical uncertainties. Currently, they are obtained
from lightcone sum rules (LCSRs) \cite{Khodjamirian:2010vf, Ball:2004ye}, restricted 
to the low-$q^2$ region. At high $q^2$, ongoing efforts aim at the first unquenched 
predictions from the lattice \cite{Bouchard:2012tb}. Current predictions of observables
at high $q^2$ rely on extrapolations of the low-$q^2$ LCSR results.

In the absence of precise form factor predictions, it is still possible to constrain
new physics with the help of ``optimized'' observables, i.e., observables that exhibit
reduced sensitivity to the form factors. On the experimental side, this requires
an angular analysis of the 4-body final state $B\to K^*(\to K\pi) 
\,\ell^+\ell^-$ giving access to 12 angular observables $J_i$ that are in principle
independent. Based on the form factor symmetries at low and high $q^2$, a number
of form factor insensitive combinations ($A_T^{(2,3,4,5,{\rm re}, {\rm im})}$
\cite{Kruger:2005ep}, $P_{(1,\ldots, 6)}$ \cite{Matias:2012xw} at low $q^2$ and 
$H_T^{(1,\ldots,5)}$ \cite{Bobeth:2010wg} at high $q^2$) of the $J_i$ have been
identified, which will be hopefully measured in the future. In these
observables, subleading corrections in $1/m_b$ expansions are the main 
uncertainties~\cite{Khodjamirian:2010vf, Bobeth:2010wg}.

CP-violating effects in $b\to s$ transitions are predicted to be very small
in the SM, being proportional to the doubly Cabibbo-suppressed combination
$\mbox{Im}(V_{ub}^{} V_{us}^*)$. Here several CP asymmetries of the angular
observables $J_{5,6,8,9}$ can be extracted from un-tagged $B$ meson samples
and moreover, CP asymmetries of $J_{7,8,9}$ are not suppressed by small QCD
phases \cite{Bobeth:2008ij}. The first measurements of the CP asymmetry
$A_{\rm im}$ of $J_9$~\cite{Aaltonen:2011qs} and the rate CP asymmetry of
$B\to K^* \ell^+\ell^-$ became available \cite{Aaij:2012cq}, where the latter
equals to the one of $B\to K \ell^+\ell^-$ in the SM operator basis 
\cite{Bobeth:2011nj}.

In view of the high future accuracy of measurements of $B\to K^*(\to K\pi)
\ell^+\ell^-$, the ($K\pi$) pairs not originating from the $K^*$ decay, especially
the resonant and non-resonant $S$-wave contributions, will affect the
angular distribution. They can be controlled due to the angular analysis
and require a careful treatment on the experimental side \cite{Becirevic:2012dp}.

%
%
%

\smallskip

\begin{figure}[t]
\centerline{
\includegraphics[width=0.33\textwidth]{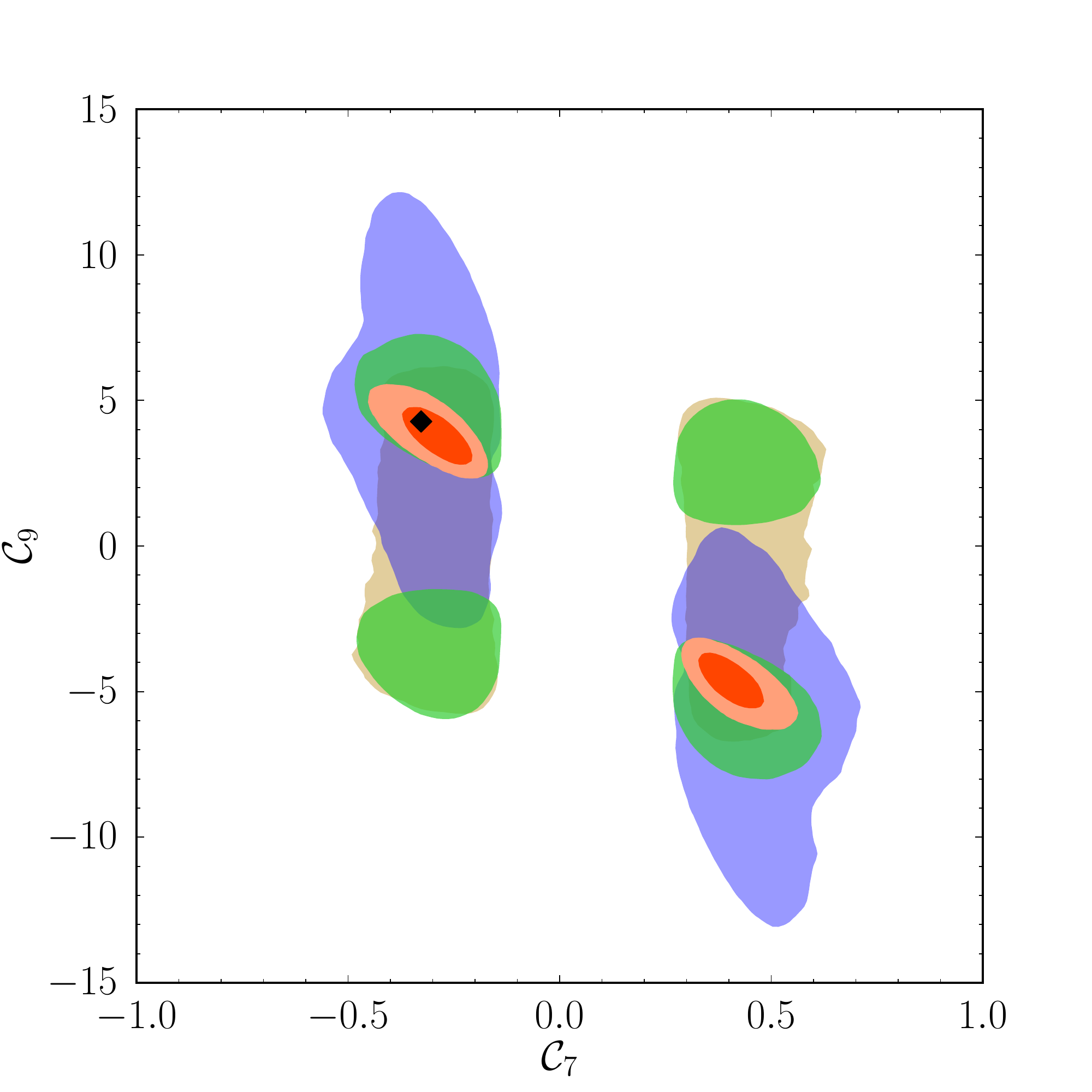}
\includegraphics[width=0.33\textwidth]{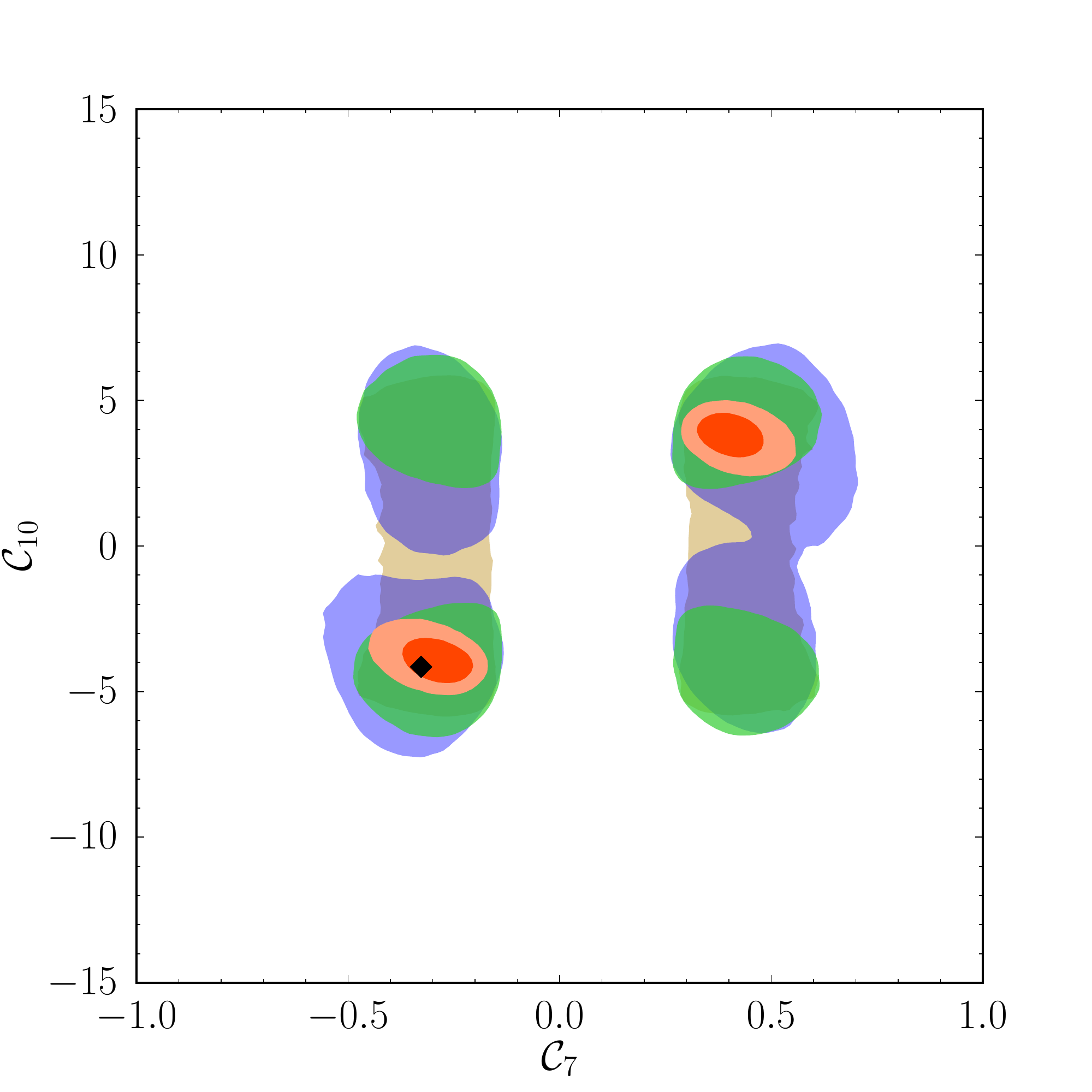}
\includegraphics[width=0.33\textwidth]{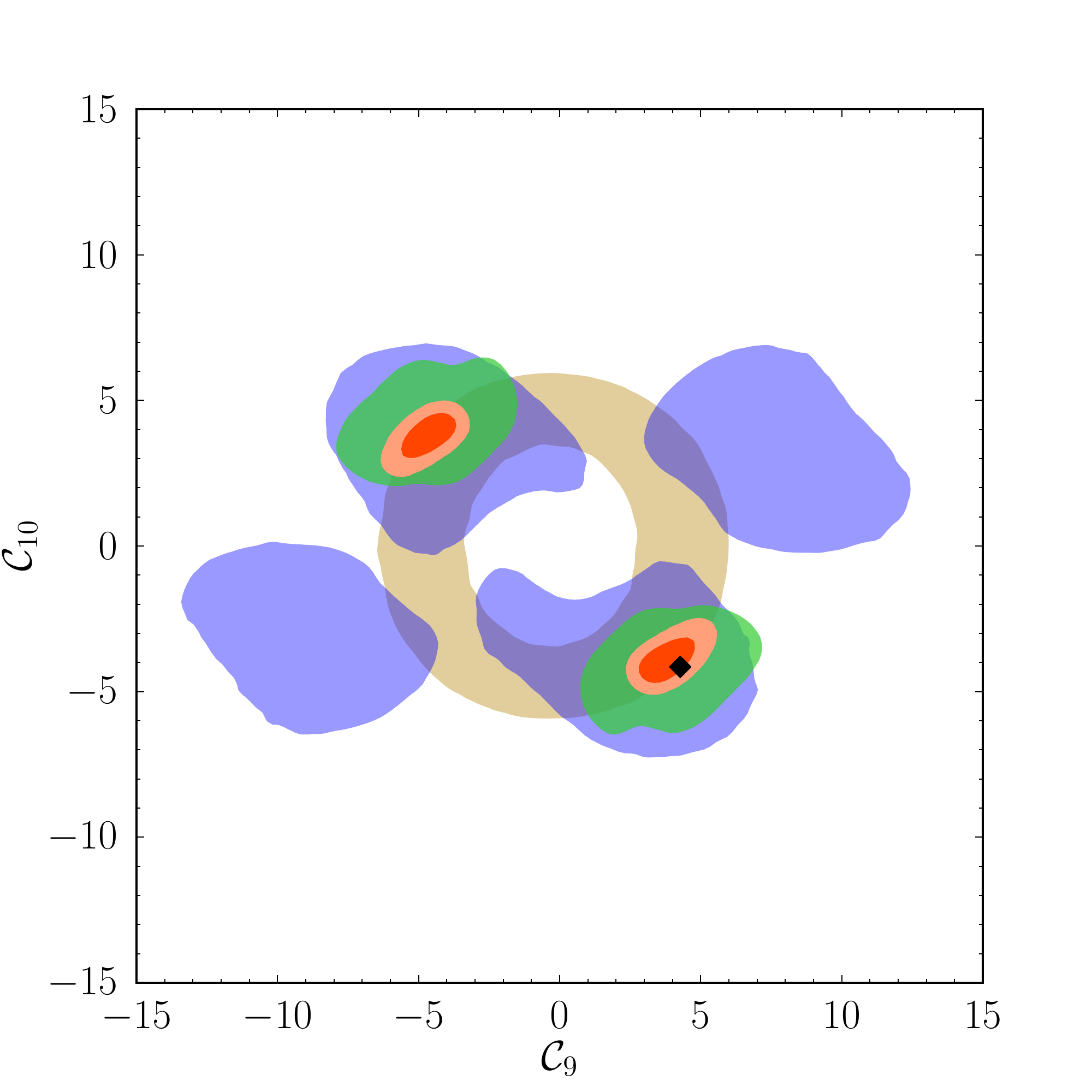}
}
\caption{The marginalized 2-dim 95\% credibility regions of ${\cal C}_{7,9,10}$ 
   at $\mu = 4.2$ GeV are shown when combining the $B\to K^* \gamma$ constraints
   with data from 
   {\it i)} only low- and high-$q^2$ $B \to K \ell^+\ell^-$ [brown]; 
   {\it ii)} only low-$q^2$ $B \to K^*\ell^+\ell^-$ [blue];
   {\it iii)} only high-$q^2$ $B \to K^*\ell^+\ell^-$ [green]; and 
   {\it iv)} all the data, including also $B_s \to \mu^+\mu^-$ [light red], 
   showing as well the 68\% credibility region [red].
   The SM values of ${\cal C}^{\rm SM}_{7,9,10}$ are indicated by $\filleddiamond$.}
\label{fig:SM-fit}
\end{figure}

{\bf Model-independent constraints:} A scenario of real\footnote{It is customary
to factor out the complex CKM combination $V_{tb}^{} V_{ts}^{*}$, i.e., implying
minimal flavor violation.} nonstandard contributions $C_{7,9,10}$ can be considered
as the simplest model-independent extension of the Standard Model (SM). Here
$C_i$ denote the short-distance couplings (Wilson coefficients) of the numerically
most important $b\to s \gamma$
and $b\to s\,\ell^+\ell^-$ mediating operators in the $|\Delta B| = |\Delta S| = 1$
effective Hamiltonian known in the SM at the next-to-next-to leading order
\cite{Bobeth:1999mk}. Combining available results\footnote{
The following measurements were not included: {\it i)} the CDF final
data set with 9.6 fb$^{-1}$ presented at the ICHEP-2012 conference 
\cite{Aaltonen:2011qs}, {\it ii)} the recent world-best measurements of 
$B\to K\,\ell^+\ell^-$ from 1~fb$^{-1}$ of LHCb \cite{Aaij:2012cq} and {\it iii)}
first evidence of $B_s \to \mu^+\mu^-$ by LHCb \cite{:2012ct}.} of the exclusive 
channels $B\to K^*\gamma$, $B\to K^{(*)}\ell^+\ell^-$ and $B_s\to \mu^+\mu^-$, 
one obtains the 2-dim marginalized posterior distributions as shown in figure 
\ref{fig:SM-fit} \cite{Beaujean:2012uj}. Theory uncertainties have been included
as nuisance parameters and are marginalized over, the fit also provides updated
knowledge on them. Two solutions remain viable, one including the SM and the
other with all sign-flipped Wilson coefficients. The goodness-of-fit yields
satisfactory $p$ values between 0.60 and 0.75 for both solutions. Also the SM 
indicates a good fit \cite{Altmannshofer:2011gn}. Observables
sensitive to 4-quark operator contributions, such as ${\cal B}(B\to X_s \gamma)$, 
$A_{\rm FB}(B\to K^*\ell^+\ell^-)$ or $A_I(B\to K^*\gamma)$, provide means to 
distinguish the two solutions \cite{Matias:2012xw}.

The experimental results also imply constraints on scenarios beyond the minimal
setup, such as allowing for CP violation beyond the standard CKM picture 
\cite{Altmannshofer:2011gn, Bobeth:2011nj} and including additional operators:
chirality-flipped $C_{7',9',10'}$ and/or (pseudo)-scalar $C_{S,S', P, P'}$ 
\cite{Altmannshofer:2011gn, Becirevic:2012fy, Bobeth:2011st}.
The data yield strong correlations among the allowed regions of the Wilson 
coefficients, including their moduli and phases, which might not be easily
visualized beyond 2-dim marginalized plots. Consequently, predictions for not yet
measured observables in theses scenarios are more informative. The data still
allow for sizable CP-violation \cite{Altmannshofer:2011gn} for the CP
asymmetries $A_{7,8,{\rm im}}$. Moreover, right-handed currents start to
be constrained from the interplay of branching-ratio measurements of 
$B\to K^*\ell^+\ell^-$ and $B\to K\ell^+\ell^-$ \cite{Altmannshofer:2011gn}.

The decays $B\to K\,\ell^+\ell^-$ and $B_s\to \mu^+\mu^-$ provide complementary 
information on chirality-flipped Wilson coefficients which enter both modes
as $(C_i + C_{i'})$ and $(C_i - C_{i'})$, respectively. Their interplay has
been explored in some detail for $C_{S,P,10}$ in \cite{Becirevic:2012fy}. The
decay $B\to K\,\ell^+\ell^-$ allows also to constrain (pseudo-)scalar and tensor
operators with the help of the observables $A_{\rm FB}$ and $F_H$ that are
accessible in the angular analysis \cite{Beneke:2001at}. In this respect,
latest data from LHCb on $F_H$ at high $q^2$ \cite{Aaij:2012cq} provide updated
constraints on $|C_T|^2 + |C_{T5}|^2 \lesssim 0.5$ \cite{Bobeth:2010wg}.

The transition $b\to s\,\tau^+\tau^-$ is experimentally not constrained except for
the upper bound ${\cal B}(B^+ \to K^+\tau^+\tau^-) < 3.3\cdot 10^{-3}$ from BaBar
\cite{Flood:2010zz}. Due to mixing of $b\to s\,\tau^+\tau^-$ operators into
$b\to s\,(\gamma,\, \ell^+\ell^-)$ ($\ell = e,\, \mu$), the latter processes imply
indirect constraints on large nonstandard $b\to s\,\tau^+\tau^-$ contributions
\cite{Bobeth:2011st}. When combined with the direct constraint on ${\cal B}(B^+
\to K^+\tau^+\tau^-)$, they rule out an enhancement of the
decay width difference $\Delta \Gamma_s$ of the $B_s$-meson of more than $35$\%
compared to the SM prediction, assuming single operator dominance~\cite{Bobeth:2011st}. 

At the moment, the measured observables in exclusive $b\to s\, \ell^+\ell^-$ decays
can be explained within the SM and they push the scale of tree-level-mediated 
new FCNC interactions in this sector above ${\cal O}(20\, \mbox{TeV})$ -- for some
even above ${\cal O}(100\, \mbox{TeV})$ -- assuming single operator dominance and
order-one couplings \cite{Bobeth:2011nj, Beaujean:2012uj, Altmannshofer:2011gn}. 
The measurement of additional observables, especially angular observables in 
$B\to K^*(\to K\pi)\, \ell^+\ell^-$, will allow to further scrutinize nonstandard
interactions. The absence of any new-physics signal in the future, within uncertainties,
will put strong constraints on model parameter spaces in model-dependent analyses.

%
%
%

\smallskip

{\bf Model-dependent constraints:} In the past, the most frequently considered 
$|\Delta B| = |\Delta S| = 1$ FCNC observables were ${\cal B}(B\to X_s \gamma)$
and the upper bound on ${\cal B}(B_s \to \mu^+\mu^-)$ in order to place constraints
on parameter spaces of extensions of the SM, especially for the supersymmetric ones
(MSSM, NMSSM, $\ldots$). The $b\to s\gamma$ mediated decays test especially 
chirality-changing effects, whereas $B_s \to \mu^+\mu^-$ is very sensitive to
(pseudo-) scalar contributions as for example neutral Higgs penguins. To less
extent, analyses of tree-level FCNC's have resorted to $B\to X_s \ell^+\ell^-$. 
However, the new data on exclusive $B\to K^{(*)} \ell^+\ell^-$ decays allow
also to test flavor-changing $Z$ and $Z'$-couplings to $b$ and $s$ quarks as
well as modifications of left- and right-handed $W$ couplings that are present
in many extensions of the SM. In the effective theory they modify the Wilson
coefficients $C_{7,9,10}$ and their chirality-flipped counterparts $C_{7',9',10'}$.

In the framework of the MSSM, the latest $b\to s\, \ell^+\ell^-$ data constrain 
flavor-changing left-right mixing $(\delta^u_{23})_{LR}$ in the up-squark sector,
which in turn places constraints on various FCNC decays $t\to c\gamma$, $t\to cg$
and $t\to cZ$ of the top quark \cite{Behring:2012mv}.
The interplay of $B_s\to \mu^+\mu^-$ at large $\tan\beta$ and angular observables
in $B\to K^* \ell^+\ell^-$ at moderate $\tan\beta$ has been investigated in
constrained scenarios such as the CMSSM and NUHM~\cite{Mahmoudi:2012un}.

LeptoQuark interactions, which induce scalar and pseudo-scalar operators 
${\cal O}_{S,S',P,P'}$, have been constrained with recent data from $B\to (X_s,
K)\, \ell^+\ell^-$, and $B_s\to \mu^+\mu^-$ \cite{Kosnik:2012dj}.

The latest data of exclusive $b\to s\,\ell^+\ell^-$ decays provide also constraints
on models with extended gauge-sectors $Z'$ \cite{Buras:2012jb}, especially
beyond minimal flavor-violating scenarios, but also in some models of partial 
compositeness where they probe interactions with $Z$ bosons \cite{Barbieri:2012tu}. 

\smallskip

{\bf Acknowledgments:} We are indebted to the organizers of the CKM 2012 
for the opportunity to present a talk and the kind hospitality in Cincinnati.

%
%

\end{document}